\documentclass[preprint,12pt]{elsarticle}

\usepackage{amssymb}

\journal{Chinese Journal of Physics}

\begin{document}

\begin{frontmatter}

\title{Behaviour of pf shell under RMF+BCS Description}

\author[a]{G. Saxena} \author[b]{M. Kaushik}
\address[a]{Department of Physics, Govt. Women Engineering College, Ajmer-305002, India}
\address[b]{Department of Physics, Shankara Institute of
Technology, Kukas, Jaipur-302028, India}



\begin{abstract}
We have employed RMF+BCS (relativistic mean-field plus BCS)
approach to study behaviour of pf shell with the help of ground
state properties of even-even nuclei. Our present investigations
include separation energies , deformations, single particle energies, wavefunction, potential as well
etc density distribution. As per recent experiments showing neutron magicity at N = 32 for Ca isotopes,
our results with mass dependent pairing indicate a shell closure at N = 32
in Ca isotopes and a more strong shell closure at N = 34 in proton
deficient $^{48}$Si because of reorganization of neutron pf shell. In a
similar manner, proton pf shell structure is more likely to
produce shell closure at Z = 34 with a doubly magic character for
$^{84,116}$Se. We have also included N = 40 isotones and Z = 40 isotopes for our
study and predicted $^{60}$Ca and $^{68}$Ni as doubly magic nuclei out of which $^{60}$Ca is found near drip-line of Ca and a potential candidate for
future studies in the chain of Ca isotopes next to doubly magic $^{52}$Ca.
\end{abstract}


\end{frontmatter}


\section{Introduction}
Evolution of shell structure is being a momentous topic in the world of
theoretical and experimental nuclear physics since last two decades.
There are several regions of periodic chart which have been
established showing new magicity or showing disappearance of
traditional magicity through different theoretical approaches and
confirmed from many experiments. As an example, appearance of the new magic
number N = 16 \cite{ozawa,kanungo,hoffman} has been corroborated and neutron
drip-line nucleus $^{24}$O with N = 16, is now
considered as a doubly-closed-shell nucleus \cite{kanungo,hoffman,tshoo}.
On the other hand disappearance of the conventional magic numbers N = 8, 20
and 28 have been demonstrated through many communications
\cite{iwasaki} - \cite{otsuka}.

In recent past, experimental studies strongly indicate N = 32 as a new
magic number in Ca isotopes due to the high energy of the first 2$^+$
state in this nucleus \cite{gade}. In addition, high precision mass
measurements were performed for the neutron rich Ca isotopes
$^{53}$Ca and $^{54}$Ca by employing the mass spectrometer of
ISOLTRAP at CERN and confirmed the magicity of the nucleus $^{52}$Ca
\cite{wienholtz}. Furthermore, first experimental spectroscopic study on
low–lying states was performed also with proton knockout
reactions at RIKEN indicating the magic nature of the nucleus
$^{54}$Ca \cite{stepp}. More recently, in 2015, N = 32 shell closure
is confirmed for exotic isotopes $^{52,53}$K in Ref. \cite{rosenbusch}.

These recent evidences for emerging sub-shell gaps at neutron number
N = 32 and N = 34 have accumulated high interest for the description of pf-shell nuclei
\cite{gallant,jia,grasso-ca}. The next benchmark in this region of pf shell after N = 32 and N = 34
is N = 40 \cite{wang1}, which is important in determining nuclear structure and likely also the location of
the neutron drip-line in the Ca isotopic chain \cite{meng}.

The theoretical studies of such nuclei can be in
general grouped into three different approaches. These are (i) the ab initio
methods, (ii) the macroscopic models with shell corrections and (iii)
the self-consistent mean-field and shell model theories. As far as mean field models are concern there are
mainly three nuclear mean-field models which are widely used in current calculations and
are based on (i) the Skyrme zero range interaction initially employed
by Vautherin and Veneroni \cite{vautherin1}, and Vautherin and Brink \cite{vautherin}, (ii) the
Gogny force \cite{gogny} with finite range, and (iii) the relativistic mean-field
model formulated by Walecka \cite{walecka}, and Boguta and Bodmer \cite{boguta}. During the last several years,
theoretical studies of neutron and proton rich nuclei away from the valley of $\beta$-stability have been mostly
accomplished under the framework of mean-field theories \cite{terasaki,dobac00,grasso,sand2}, and
also employing their relativistic counterparts \cite{walecka,boguta,bouy,pgr2,gambhir,suga,ring,sharma,meng4,lala,mizu,estal,yadav,yadav1,meng2}.
Recently, the interest in tensor interaction for nuclear structure has also been
revived by the study of the evolution of nuclear properties far from
the stability line \cite{jia,utsuno,grasso2}.
However, in the context of the effective theories that describe medium-heavy nuclei, the role of the tensor
force is still debated \cite{sagawa}.

In this paper, we will target pf shell nuclei with N(Z) = 32 and 34 along with the nuclei with N(Z) = 40 to study shell evolution for complete
chain of isotones(isotopes) upto drip-lines (covering mass region from A = 48 to A = 122). When nuclei move away from stability line toward their drip-lines then corresponding Fermi surface moves closer to zero energy at the continuum threshold and therefore significant number of the available single-particle states form part of the continuum. Since pairing correlations are very important for drip-line nuclei therefore to include them together with a realistic mean-field, the HF+BCS and RMF+BCS calculations has turned out to be very useful and successful tool as has been demonstrated recently \cite{saxena,saxena1}. The principal
advantage of the RMF approach is that it provides the spin-orbit interaction in the entire mass region in a natural way. Since the single particle properties near the threshold are prone to large changes as compared to the case of deeply bound levels in the nuclear potential therefore RMF approach has proved to be very crucial for the study of unstable nuclei upto the drip-lines. The results provided by RMF+BCS scheme \cite{saxena1,singh} are indeed in close agreement with the experimental data and other similar mean-field calculations \cite{meng}. In our investigations of pf shell nuclei, we use RMF+BCS approach to calculate ground state properties viz. single particle energy, deformation, separation energy as well as pairing energy etc. The results are compared with recent experimental results and various popular force parameters viz. TMA \cite{suga-tma}, NLSH \cite{ring3-nlsh}, NL3 \cite{lalanl3}, PK1 \cite{pk1} and NL3* \cite{nl3star} to bear witness of our results and outcomes. These above sets of parameters are for non-linear interaction and have been commonly used for variety of mass calculations for whole periodic region \cite{ring,sharma,yadav,singh,geng1} with a very good agreement with available experimental data for whole mass table. Meanwhile, the new parameter sets belong to density dependent coupling are DD-ME1 \cite{nik}, DD-ME2 \cite{lalaDD}, DD-ME$\delta$ \cite{roca} which have also provided a more realistic description of neutron matter and finite nuclei. However, due to few limitations of these density dependent coupling in describing several transitional medium-heavy nuclei \cite{ring-new} and to make a simplified and general treatment, we here prefer non-linear coupling to describe our results of pf shell nuclei which may posses transitional character \cite{takeuchi,heyde,otsuka-new}. The non linear interaction in RMF theory has proven to be very much effective, consistent and reliable \cite{yadav,yadav1,meng2,saxena,saxena1,singh} and has recently successfully applied to study deformed nuclei at proton drip line \cite{lidia}, in the study of neutron star \cite{xing}, in describing properties of superheavy nuclei \cite{mehta}  and determining hadron-quark phase transition line in the QCD phase diagram \cite{junpei}. Therefore, we strongly believe that conclusions drawn here using RMF+BCS calculations with non linear effective interaction will not be affected very much by other interactions or approaches.

\section{Relativistic Mean-Field Model}
The calculations have been carried out using the following model
Lagrangian density with nonlinear terms both for the ${\sigma}$ and
${\omega}$ mesons as described in detail in Refs.~\cite{suga,yadav,singh}.

\begin{eqnarray}
       {\cal L}& = &{\bar\psi} [\imath \gamma^{\mu}\partial_{\mu}
                  - M]\psi\nonumber\\
                  &&+ \frac{1}{2}\, \partial_{\mu}\sigma\partial^{\mu}\sigma
                - \frac{1}{2}m_{\sigma}^{2}\sigma^2- \frac{1}{3}g_{2}\sigma
                 ^{3} - \frac{1}{4}g_{3}\sigma^{4} -g_{\sigma}
                 {\bar\psi}  \sigma  \psi\nonumber\\
                &&-\frac{1}{4}H_{\mu \nu}H^{\mu \nu} + \frac{1}{2}m_{\omega}
                   ^{2}\omega_{\mu}\omega^{\mu} + \frac{1}{4} c_{3}
                  (\omega_{\mu} \omega^{\mu})^{2}
                   - g_{\omega}{\bar\psi} \gamma^{\mu}\psi
                  \omega_{\mu}\nonumber\\
               &&-\frac{1}{4}G_{\mu \nu}^{a}G^{a\mu \nu}
                  + \frac{1}{2}m_{\rho}
                  ^{2}\rho_{\mu}^{a}\rho^{a\mu}
                   - g_{\rho}{\bar\psi} \gamma_{\mu}\tau^{a}\psi
                  \rho^{\mu a}\nonumber\nonumber\\
                &&-\frac{1}{4}F_{\mu \nu}F^{\mu \nu}
                  - e{\bar\psi} \gamma_{\mu} \frac{(1-\tau_{3})}
                  {2} A^{\mu} \psi\,\,,
\end{eqnarray}
In the above Lagrangian the field tensors $H$, $G$ and $F$ for the vector fields are
defined by
\begin{eqnarray}
                 H_{\mu \nu} &=& \partial_{\mu} \omega_{\nu} -
                       \partial_{\nu} \omega_{\mu}\nonumber\\
                 G_{\mu \nu}^{a} &=& \partial_{\mu} \rho_{\nu}^{a} -
                       \partial_{\nu} \rho_{\mu}^{a}
                     -2 g_{\rho}\,\epsilon^{abc} \rho_{\mu}^{b}
                    \rho_{\nu}^{c} \nonumber\\
                  F_{\mu \nu} &=& \partial_{\mu} A_{\nu} -
                       \partial_{\nu} A_{\mu}\,\,\nonumber\
\end{eqnarray}

and other symbols have their usual meaning.\\

Now, one can apply the BCS approximation to the state-independent pairing force but high density of single-particle
states in the particle continuum immediately results in an unrealistic increase of BCS pairing
correlations \cite{nazar}. To avoid such an increase, one may artificially readjust
the pairing strength constant but then the spatial asymptotic properties of the solutions become incorrect and consequently due to a nonzero occupation probability of quasibound states, there appears an unphysical gas of neutrons surrounding the nucleus \cite{dobaczewski}. These deficiencies can be healed by applying state-dependent-pairing-gap version, where the pairing gap is calculated for every single-particle state. But it is found that even with state dependent BCS calculations a surplus density above asymptotic limit appears at large distances results incorrect behavior of the density \cite{dobaczewski}. Therefore, one may think that excluding the scattering states from the pairing phase space could be a decisive solution to the problem. Indeed, such scheme has been introduced by Sandulescu et al. \cite{sandulescu} in which the effect of the resonant continuum on pairing correlations is introduced through the scattering wave functions located in the region of the resonant states. These states are found by solving the relativistic mean field equations with scattering-type boundary conditions for continuum spectrum and this scheme rBCS is found very effective for the description of drip line nuclei \cite{sandulescu}.

In this paper, for the chain of Ca and Ni isotopes, we have found resonant states 1g$_{9/2}$ and 1g$_{7/2}$ respectively but these states start to play an important role with their resonant part for N $>$ 40 and N $>$ 50 respectively. With this in view, these contributions of resonant states are less significant in terms of discussion of pf shells considered in this paper where we are concern upto N = 40 only. Moreover, we are mainly dealing with doubly magic nuclei in this paper and for doubly magic nuclei results of RMF+BCS and RMF+rBCS (including resonant part by Sandulescu et al. \cite{sandulescu}) are found exactly same because for doubly magic nucleus pairing energy vanishes and resonant state does not contribute in the pairing energy. Therefore, in the context of this paper where we target only for pf shell nuclei (N or Z $\leq$ 40), it is reasonable to perform state dependent BCS calculations \cite{lane,ring2} based on single-particle spectrum in which continuum is replaced by a set of positive energy states generated by enclosing the nucleus in a spherical box. We have chosen box radius R = 30 fm which is same as has been taken in Ref \cite{yadav,yadav1,saxena,saxena1} and results are not much affected by changing the box radius around R = 30 fm at least for the considered doubly magic nuclei in pf shell.

Thus the gap equations have the standard form for all the single
particle states, i.e.
\begin{eqnarray}
     \Delta_{j_1}=-\frac{1}{2}\frac{1}{\sqrt{2j_1+1}}
     \sum_{j_2}\frac{\left<{({j_1}^2)\,0^+\,|V|\,({j_2}^2)\,0^+}\right>}
      {\sqrt{\big(\varepsilon_{j_2}\,-\,\lambda \big)
       ^2\,+\,{\Delta_{j_2}^2}}}\,\,\sqrt{2j_2+1}\,\,\, \Delta_{j_2}\,\,
\end{eqnarray}\\
where $\varepsilon_{j_2}$ are the single particle energies, and
$\lambda$ is the Fermi energy, whereas the particle number condition
is given by $\sum_j \, (2j+1) v^2_{j}\, = \,{\rm N}$. In the
calculations we use for the pairing interaction a delta force, i.e.,
V = -V$_0 \delta(r)$. Initially, we use the same value of V$_0$ for our present study (V$_0$ = 350 MeV
fm$^3$) which was determined in Ref.~\cite{yadav1} by obtaining a best
fit to the  binding energy of Ni isotopes. Since, we are covering region from A = 48 to A = 122 therefore we will examine our results using mass dependent ($1/A$-dependancy) pairing strength which may be an effective way to handle large mass region. Apart from its simplicity, the applicability and justification of using
such a $\delta$-function form of interaction has been discussed in Ref. \cite{dobac-delta}, whereby it has been shown in the
context of HFB calculations that the use of a delta force in a
finite space simulates the effect of finite range interaction in a
phenomenological manner (see also \cite{bertsch91}
for more details). The pairing matrix element for the
$\delta$-function force is given by\
\begin{eqnarray}
\left<{({j_1}^2)\,0^+\,|V|\,({j_2}^2)\,0^+}\right>
=-\,\frac{V_0}{8\pi}
       \sqrt{(2j_1+1)(2j_2+1)}\,\,I_R\,\,
\end{eqnarray}
Here V$_0$ is the pairing strength which is taken as 350 MeV
fm$^3$ for initial calculations and then it is used with mass dependency for further calculations as mentioned above. Mass dependent pairing strength used here is given by
\begin{eqnarray}
V_0 = \frac{25,500}{A} MeV fm^3
       \end{eqnarray}

In equation (3), $I_R$ is the radial integral having the form
\begin{eqnarray}
   I_R& =&\,\int\,dr \frac{1}{r^2}\,\left(G^\star_{j_ 1}\, G_{j_2}\,+\,
     F^\star_{j_ 1}\, F_{j_2}\right)^2
\end{eqnarray}
Here $G_{\alpha}$ and $F_{\alpha}$ denote the radial wave functions
for the upper and lower components, respectively, of the nucleon
wave function expressed as
\begin{equation}\psi_\alpha={1 \over r} \,\, \left({i \,\,\, G_\alpha \,\,\,
 {\mathcal Y}_{j_\alpha l_\alpha m_\alpha}
\atop{F_\alpha \, {\sigma} \cdot \hat{r}\, \, {\mathcal Y}_{j_\alpha
l_\alpha m_\alpha}}} \right)\,\,,
\end{equation}
and satisfy the normalization condition
 \begin{eqnarray}
         \int dr\, {\{|G_{\alpha}|^2\,+\,|F_{\alpha}|^2}\}\,=\,1
 \end{eqnarray}

In Eq. (6) the symbol ${\mathcal Y}_{jlm}$ has been used for the
standard spinor spherical harmonics with the phase $i^l$. The
coupled field equations obtained from the Lagrangian density in (1)
are finally reduced to a set of simple radial equations which are solved self consistently along with the equations  for the
state dependent pairing gap $\Delta_{j}$ and the total particle
number N for a given nucleus.\\

The relativistic mean field description has been extended for the
deformed nuclei of axially symmetric shapes by Gambhir, Ring and
their collaborators \cite{gambhir} using an expansion method.
The scalar, vector, isovector and charge densities, as in the
spherical case, are expressed in terms of the spinor $\pi_i$, its
conjugate $\pi_i^+$, operator $\tau_3$ etc. These densities serve as
sources for the fields $\phi$ = $\sigma$, $\omega^0$ $\rho^0$ and
$A^0$, which are determined by the Klein-Gordon equation in
cylindrical coordinates. Thus a set of coupled equations, namely the
Dirac equation with potential terms for the nucleons and the
Klein-Gordon type equations with sources for the mesons and the
photon is obtained. These equations are solved self consistently.
For this purpose, as described above, the well-tested basis
expansion method has been employed \cite{gambhir,geng1}.
For further details of these formulations we refer the
reader to refs.~ \cite{gambhir,singh,geng1}.

\section{Results and Discussion}
\begin{figure}[htbp]
\begin{center}
\includegraphics[width=110mm]{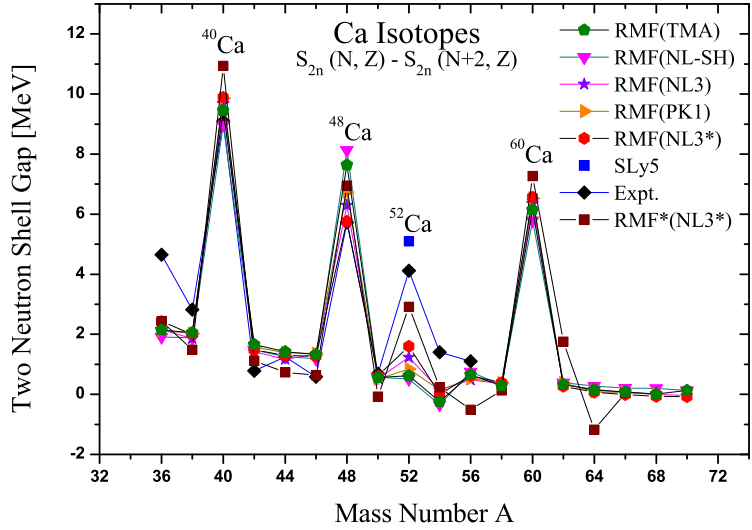}
\caption{(Colour online) Two neutron shell gap [S$_{2n}$(N, Z) - S$_{2n}$(N+2, Z)] for Ca Isotopes calculated through various parameters.
For comparison experimental data \cite{audi} and the data calculated by tensor interaction (SLy5)
\cite{grasso-ca} are also shown. RMF* is representing RMF calculations done using mass dependent pairing.} \label{fig1}
\end{center}
\end{figure}

Magic numbers between N(Z) = 20 to N(Z) = 50 are mainly governed by position and separation of single particle states
in and from $pf$ shell. For a detailed study in this region towards neutron and proton side, we first carry out RMF+BCS
calculations including the deformation degree of freedom (referred
to throughout as deformed RMF) to search spherical or nearly spherical isotones/isotopes. It is
found that some isotones/isotopes actually exhibit very small or almost no deformation.
Therefore, for such cases of negligible deformation, we take advantage to use RMF+BCS approach for spherical
shapes (referred to throughout as RMF) for the analysis of results. This description in terms of spherical single particle wave functions and energy
levels treats shell closures and magicity etc. with more convenience and transparency. Addition to this, with the spherical framework of RMF pairing gaps, total pairing energy, contributions of neutron and proton single particle states etc. can be demonstrated with more clarity.

To study pf shell, we first turn on our discussion with emergence of new shell closures (N = 32 and 34) in exotic Ca isotopes which have been
indicated by recent experiments \cite{gade,wienholtz,stepp,rosenbusch}. First it was observed that due to high excitation energy of $^{52}$Ca as compared to its neighbouring nuclei, N = 32 shell closure may develop \cite{pris,dinca}.
After that from recent high-precision mass measurements of several
isotopes ranging from $^{51}$Ca up to $^{54}$Ca a confirmation was
obtained in 2013 \cite{wienholtz}. In addition, from the more recent
measurement of the 2$_{1}$+  energy in $^{54}$Ca, it is also found that
$^{54}$Ca could be magic as well \cite{stepp}.

To investigate the possibility for $^{52}$Ca and $^{54}$Ca, we have displayed two
neutron shell gap [S$_{2n}$(N, Z) - S$_{2n}$(N+2, Z)] in Fig. 1 which is determined by calculations of RMF with different sets of
parameters i.e. TMA \cite{suga-tma}, NLSH \cite{ring3-nlsh}, NL3 \cite{lalanl3}, PK1 \cite{pk1} and NL3* \cite{nl3star}. Ca isotopes have been examined with TMA and NL-SH parameters in Ref \cite{yadav} in detail where constant pairing strength V$_0$ = 350 MeV fm$^3$ has been used. This constant pairing strength V$_0$ = 350 MeV fm$^3$ has been also used in the description of other magic nuclei \cite{saxena,saxena1}. We have used the same pairing strength for other parameters also and the results are shown together with available experimental data \cite{audi} and the available data of calculations recently done by Skyrme parameter set with tensor interaction (SLy5) \cite{grasso-ca} for comparison.

It is reflected from the Fig. 1 that calculations done in Ref \cite{yadav} with TMA \& NLSH parameter and the calculations with other
parameters retrace well with the experimental data throughout for
all isotopes of Ca, however for $^{52}$Ca and $^{54}$Ca these results are not reproducing experimental peaks in a good agreement. As can be seen from Fig. 1, for Ca isotopes the results of TMA \& NLSH parameters \cite{yadav} along with NL3 \& PK1 parameter are far enough with the experimental results. However, NL3* is showing better results but even these results for $^{52}$Ca and $^{54}$Ca are not in a good agreement with experimental data. As mentioned above, we apply mass dependent pairing also for our calculations and we use NL3* parameter which is found with better agreement comparative to other parameters. It is indeed gratifying to note that mass dependent pairing together with NL3* parameter considerably improves result of $^{52}$Ca as can be seen from Fig. 1. These results for complete chain of Ca isotopes are mentioned with RMF* in the diagram shown by magenta square.

Obtaining better peak in Fig. 1 for $^{52}$Ca with RMF*(NL3*), one can also observed sudden peaks for known other doubly magic nuclei
viz. $^{40}$Ca and $^{48}$Ca.  In addition to this, from Fig. 1, no indication is found for magic character of $^{54}$Ca. Moving further, there is again a
large peak for $^{60}$Ca much stronger than $^{52}$Ca and similar to the peaks obtained in $^{40}$Ca and
for $^{48}$Ca giving indication of more strong doubly
magic nucleus $^{60}$Ca. With the recent experiments on Ca isotopes
\cite{wienholtz,stepp} next doubly magic nucleus $^{60}$Ca is
predicted here for future experiments which may be an interesting and remarkable finding due to its doubly magicity
and its position near neutron drip-line of Ca. The above results for $^{52,54,60}$Ca are further supported by pairing energy contribution from protons and neutrons, separation energies and wavefunction characteristics.

From the above discussion it would be important to investigate neutron shell closures N = 32, 34 and 40 in detail for full isotonic chains upto their driplines. In addition from the isospin symmetry considerations \cite{otsuka}, it would be also useful to study proton shell closures Z = 32, 34 and 40 for full isotopic chains upto their driplines.  Therefore, in the following we will examine these isotonic/isotopic chain in detail using RMF approach with mass dependent pairing using NL3* parameter which has produced better outcome over constant pairing strength and various other parameters as shown above in the case of $^{52}$Ca.

\subsection{N = 32 and 34}
It is very interesting example and situation for various theoretical and
experimental studies to study N = 32 and N = 34 shell closure
simultaneously, therefore in the following discussion we will examine
N = 32 and N = 34 within the framework of RMF theory using nuclei $^{52}$Ca and $^{54}$Ca.

To show the magicity of N = 32 or N = 34, we choose some spherical or
nearly spherical nuclei in the isotopic chain of Si, Ca and Ni. It
has been shown through various studies \cite{meng,wang2,grawe} and we have also found using deformed RMF approach \cite{saxena1,singh} that expect a few cases all the nuclei in Ca and Ni isotopic chain are spherical or nearly spherical in nature. Conversely, many of the Si isotopes
are found strongly deformed through various experimental and theoretical studies \cite{jurado,takeuchi}. However, $^{22,34,48}$Si
are found to have spherical shapes which has been shown already using deformed RMF approach \cite{saxena}. From deformed RMF approach
it is found that $^{42}$Si exhibits shape coexistence with an oblate ground state being the deformation parameter $\beta_{2}$ = - 0.37 \cite{saxena,lalazissis}.
Moving towards neutron rich side for $^{44,46,48}$Si
the second minimum becomes less pronounced gradually and for drip-line isotope $^{48}$Si only one minima contributes indicating completely spherical ground
state configuration. This comment, allow us to choose $^{46,48}$Si as nearly spherical/spherical nuclei for our study.
In the following discussion we mainly select $^{46}$Si, $^{52}$Ca and $^{60}$Ni to investigate shell closure at N = 32, and $^{48}$Si, $^{54}$Ca and $^{62}$Ni for shell closure at N = 34 using spherical RMF approach.

With these above remarks, in Fig. 2, we display our results of
neutron single particle energies for isotonic chain of N = 28, 30, 32
and 34. These results are obtained using NL3* parameter \cite{nl3star} with considering mass dependency on pairing strength. We have also mentioned experimental values of single
particle energies of known doubly magic nucleus
\cite{vautherin,trache}. These values are marked with filled star in
left lower panel and it is grateful to note that our calculated
single particle energies using NL3* parameter and mass dependent pairing are indeed very close to the experimental values
specially for $^{48}$Ca which earmark us to be interpretative for the study of shell closures on the basis of single particle energy.
\begin{figure}[htbp]
\begin{center}
\includegraphics[width=125mm]{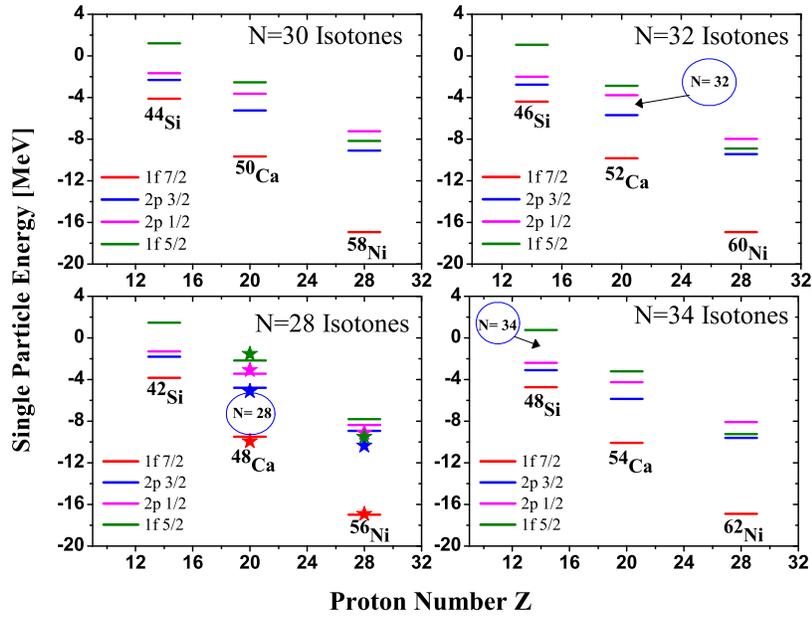}
\caption{(Colour online) Variation of neutron single particle states for Si, Ca and
Ni isotopes constituting neutron number N = 28, 30, 32 and 34 in all four panel respectively. These results are calculated using mass dependent pairing with NL3* parameter. Left lower panel
is also showing available experimental single particle energies of doubly magic
nuclei \cite{vautherin,trache}.} \label{fig2}
\end{center}
\end{figure}

The gap after neutron 1f$_{7/2}$ state is well reflected from the
figure in $^{48}$Ca and $^{56}$Ni as shown in lower left panel of Fig. 2. However this shell gap gets
weakened in $^{42}$Si suggesting a disappearance of N = 28 shell
closure similar to refs. \cite{bastin,jurado,dimitar,takeuchi}. This
weakening can be caused and inferred by inclusion of deformations
which is a matter of separate discussion in the main context of this
paper. Now moving towards more neutron rich case for N = 32 (right
upper panel), the next 4 neutrons (28 + 4 = 32 Neutrons) are filled in the states
2p$_{3/2}$ and 2p$_{1/2}$ located above the 1f$_{7/2}$ state. From
right upper panel in Fig. 2, it is evident that 2p$_{3/2}$ and
2p$_{1/2}$ states are very close to each other in $^{46}$Si, and are
occupied by these four neutrons simultaneously. However, moving
towards higher Z (from Z = 14 to Z = 20) i.e. in $^{52}$Ca, these
two states 2p$_{3/2}$ and 2p$_{1/2}$ get separated as shown
in the Fig. 2 for N = 32 isotones. This separation results a shell gap at N = 32 in $^{52}$Ca
which is similar to recent results \cite{gade,wienholtz}. It is also observed from the Fig. 2 that the state having higher angular momentum (neutron 1f$_{5/2}$)
moves deeper on increasing proton number (Si to Ni). For Si isotopes, 1f$_{5/2}$
state is far separated from closely situated 2p$_{3/2}$ and 2p$_{1/2}$ states. Whereas for Ca isotopes due to more
protons this 1f$_{5/2}$ state is quite close to 2p$_{1/2}$ state resulting
a gap after 2p$_{3/2}$ state which may even lead to N = 32 shell
closure. However, this shell gap is not so strong as it has been found for any conventional magic number such as gap after 1f$_{7/2}$ in $^{48}$Ca
as shown in left lower panel. Moving towards more proton rich side in Ni isotopes, this
state 1f$_{5/2}$ moves even below to 2p$_{3/2}$ and 2p$_{1/2}$
states and therefore this filling in of 1f$_{5/2}$ state before 2p$_{3/2}$ and
2p$_{1/2}$ states declines the possibility of N = 32 shell gap in
$^{60}$Ni or in further proton rich nuclei.

From the Fig. 2, it is also apparent that for N = 34 isotones
(right lower panel) for lower Z the state 1f$_{5/2}$ lies
significantly above from the states 2p$_{3/2}$ and 2p$_{1/2}$ as shown in the case of
$^{48}$Si and gives rise to new shell closure at N = 34 for lower Z
nuclei like Si. It can also be anticipated that due to large gap as reflected from Fig. 2, N =
34 shell closure behaves more strongly than that of N = 32 shell closure.
Consequently, $^{48}$Si may be considered as a more strong spherical doubly magic candidate for the future experiments in the same mass
region as that of $^{52}$Ca.

To check the dependency of parameters on the above results of N = 32 and N = 34 shell closures, it would be important and interesting to go through
the calculations with different sets of parameters. With this in view, in the Table 1, we have
tabulated results using few more parameters for comparison. We have
performed the RMF calculations with mass dependent pairing strength using TMA \cite{suga-tma}, NLSH
\cite{ring3-nlsh}, NL3 \cite{lalanl3}, PK1 \cite{pk1} and NL3*
\cite{nl3star} force parameters. The energy gaps between 2p$_{3/2}$
and 2p$_{1/2}$ states for N = 32 and between 2p$_{1/2}$ and
1f$_{5/2}$ states for  N = 34 are tabulated for different parameters
sets in Table 1.

\begin{table*}
\vspace{0.1cm} \centering \caption{Shell gap responsible for magicity of $^{52}$Ca and $^{48}$Si are shown using various parameters}
\bigskip
\begin{tabular}{|c|c|c|c|c|c|c|}
 \hline
 \multicolumn{1}{|c|}{Nucleus}&\multicolumn{1}{|c|}{Gap between states}&
 \multicolumn{5}{|c|}{Shell Gap [MeV]}\\
 \hline
 \multicolumn{1}{|c|}{}&
  \multicolumn{1}{|c|}{}&
    \multicolumn{5}{|c|}{RMF* (Mass Dependent Pairing)}\\
  \hline  \multicolumn{1}{|c|}{}&
   \multicolumn{1}{|c|}{}&
   \multicolumn{1}{|c|}{TMA}&
 \multicolumn{1}{|c|}{NLSH}&
 \multicolumn{1}{|c|}{NL3}&
 \multicolumn{1}{|c|}{PK1}&
 \multicolumn{1}{|c|}{NL3*}\\
 \hline
$^{52}$Ca& 2p$_{3/2}$ and 2p$_{1/2}$ &1.48& 1.59&1.85&1.58&1.90\\
\hline
$^{48}$Si& 2p$_{1/2}$ and 1f$_{5/2}$ &2.89&2.90&3.11&3.00&3.15\\

\hline
\end{tabular}
\label{tab15}
\end{table*}

From Table 1, it is clear that for $^{52}$Ca, the similar shell gap
between 2p$_{3/2}$ and 2p$_{1/2}$ states also lies for other parameters. Using mass dependent pairing with TMA \& NLSH parameter (see column 3 and 4) this gap is rather small [1.48 MeV for TMA and 1.59 MeV for NLSH] comparative to other parameters (column 5, 6, 7), but it is found larger than the gap calculated using constant pairing strength in Ref \cite{yadav}. With the calculations in Ref \cite{yadav}, this gap is found 1.37 MeV and 1.48 MeV with TMA and NL-SH parameter respectively. Therefore from here it may be concluded with the comparison that calculation considered here with mass dependent pairing provides a better treatment over constant pairing calculation as done in Ref \cite{yadav} specially for new shell closure in Ca isotopes. In a similar manner, the gap between 2p$_{1/2}$ and 1f$_{5/2}$ states is found significant by
other parameters resulting a more strong shell closure at N = 34 for $^{48}$Si. It is important to note here from the Table 1 that present results with NL3* parameter are more supportive and certificatory for other sets of parameters.

\begin{figure}[htbp]
\begin{center}
\includegraphics[width=125mm]{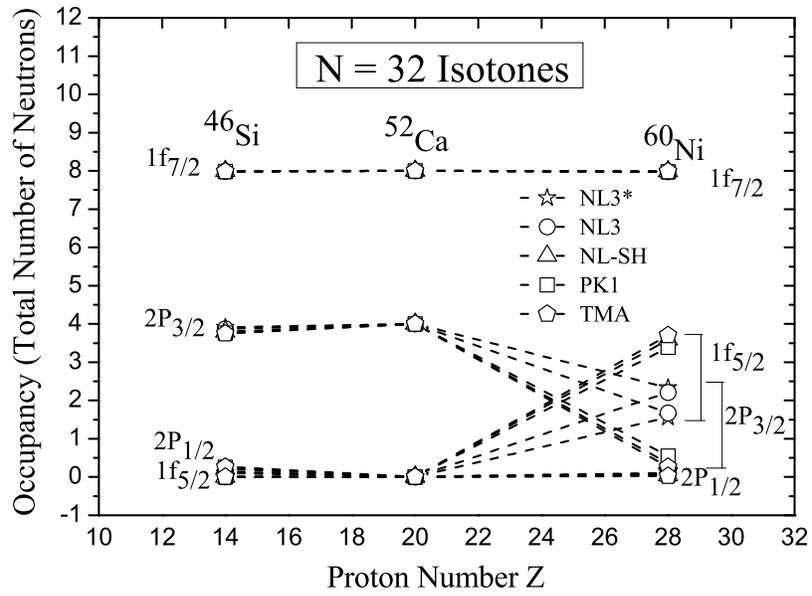}
\caption{Variation of occupancy (no. of neutrons) for various
neutron single particle states for N = 32 isotones ($^{46}$Si,
$^{52}$Ca and $^{60}$Ni).} \label{fig3}
\end{center}
\end{figure}

To get further support for above results of shell closure at $^{52}$Ca
we have displayed in Fig. 3, occupancy
(number of neutrons) of neutron single particle states 1f$_{7/2}$,
2p$_{3/2}$, 2p$_{1/2}$ and 1f$_{5/2}$ (pf shell) for the nuclei $^{46}$Si,
$^{52}$Ca, $^{60}$Ni from N = 32 isotonic chain using TMA, NL-SH, NL3, PK1 and NL3* parameters.
Very first observation from Fig. 3 is that our results from all parameters
are similar. It can be also observed that for all parameters, occupancy of state 1f$_{7/2}$ is
unchanged for all considered nuclei. However,
occupancy for other neutron states 2p$_{3/2}$, 2p$_{1/2}$ and 1f$_{5/2}$
change significantly. From the Fig. 3, it is evident that
occupancy of 1f$_{5/2}$ state for $^{46}$Si is zero as 1f$_{5/2}$ lies
around 3 MeV above than 2p$_{3/2}$ and 2p$_{1/2}$ states as
mentioned in Table 1. In $^{46}$Si these two states 2p$_{3/2}$ and
2p$_{1/2}$ are occupied together but moving towards $^{52}$Ca,
occupancy of 2p$_{1/2}$ state decreases and becomes zero. This is rather very essential for development
of N = 32 shell closure along with 2p$_{3/2}$ should attain its maximum occupancy.
Occupancy  of 2p$_{1/2}$ becomes exactly zero and occupancy of 2p$_{3/2}$ state increases upto its maximum and it accommodates all four neutrons with 2p$_{1/2}$ and 1f$_{5/2}$ vacant. The gap therefore between 2p$_{3/2}$ and 2p$_{1/2}$ leads to shell closure characteristic at N = 32. As said above, all the parameters are favourable and clearly showing magicity at N = 32 as can be seen from Fig. 3. Moving further towards $^{60}$Ni state 1f$_{5/2}$ goes much deeper due to which it accommodates more particles than 2p$_{3/2}$ and hence destroys the shell closure which has developed for
$^{52}$Ca. The calculations from TMA, NL-SH, NL3, PK1 and NL3* provide same sort
of results and behaviour of these states.

\begin{figure}[htbp]
\begin{center}
\includegraphics[width=125mm]{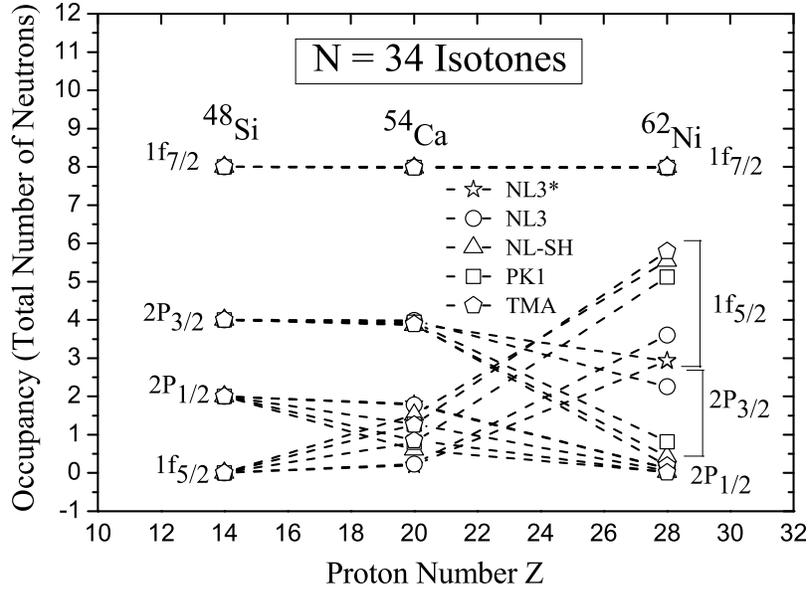}
\caption{Same as Fig. 3 but for N = 34
isotones ($^{48}$Si, $^{54}$Ca and $^{62}$Ni).} \label{fig4}
\end{center}
\end{figure}

Next in Fig. 4, we have shown occupancy
(number of neutrons) of neutron single particle pf states for the nuclei
$^{48}$Si, $^{54}$Ca, $^{62}$Ni from N = 34 isotonic chains. From Fig. 4, it can be observed that 2p$_{3/2}$ and 2p$_{1/2}$ states
are fully occupied for $^{48}$Si and 1f$_{5/2}$ is having zero
occupancy. This result is same for all parameters
and hence supporting magic character of N = 34 in $^{48}$Si. It is
worth here to point out that moving from $^{48}$Si to $^{54}$Ca
occupancy of 2p$_{3/2}$ and 2p$_{1/2}$ states decreases whereas
occupancy of 1f$_{5/2}$ state increases upto
$^{62}$Ni. Because of this filling in of 1f$_{5/2}$ state together with 2p$_{3/2}$ and 2p$_{1/2}$ states in
$^{54}$Ca there is no possibility of N = 34 shell closure in $^{54}$Ca or in
higher nuclei. However, this kind of uniform filling of 1f$_{5/2}$
state is not observed in $^{52}$Ca as illustrated above and in Fig. 3. Therefore, it can be concluded here that $^{52}$Ca and $^{48}$Si are found
to possess magic character and there are no sign found for shell closure at N = 34 in $^{54}$Ca.

\subsection{Z = 32 and 34}
After the discussion on neutron shell closure at N = 32 and 34, it
is expected from the isospin symmetry considerations \cite{otsuka}
that towards proton side Z = 32 and Z = 34 should also exhibit
similar shell closures. With this in view, in a similar manner, we
have performed our deformed RMF+BCS calculations for isotopic chain
of Z = 32 and Z = 34. Most of the nuclei in the isotopic chains are
found deformed but it is also found that there are several nuclei
which are actually spherical or nearly spherical. These are
$^{60,82,114}$Ge for Z = 32 and $^{62,84,116}$Se for Z = 34
corresponding to N = 28, 50 and 82. It is important to note that
these nuclei cover complete isotopic chain from proton rich
side to neutron rich side as one moves from N = 28 to N = 82. It is
worth to mention here that N = 82 is found to be neutron drip-line
for Ge (Z=32) and Se (Z=34) isotopic chains.

\begin{figure}[htbp]
\begin{center}
\includegraphics[width=120mm]{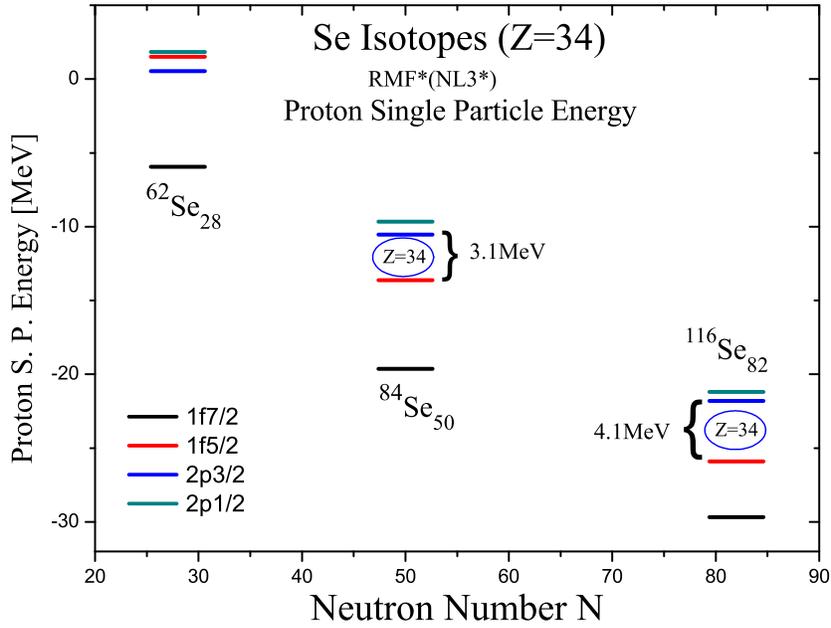}
\caption{(Colour Online) Proton single particle
energies of 1f$_{7/2}$, 1f$_{5/2}$, 2p$_{3/2}$ and 2p$_{1/2}$ states
for Se (Z=32) isotopes.} \label{fig5}
\end{center}
\end{figure}

For these nuclei, we have performed spherical RMF+BCS calculations using mass dependent pairing with NL3* parameter and
it is gratifying to note that Z = 34 indeed shows shell closure
characteristic towards neutron rich side. To elaborate this, in Fig.
5, we display our results of proton single particle energy for all above mentioned nuclei. From
Fig. 5, it is observed that proton 1f$_{5/2}$ state lies in between 2p$_{3/2}$ and 2p$_{1/2}$ states initially for proton rich side at $^{62}$Se.
Moving towards neutron rich side 1f$_{5/2}$ state goes down to these 2p$_{3/2}$ and 2p$_{1/2}$ states, consequently, creates a
significant gap (3.1 MeV for N = 50 and 4.1
MeV for N = 82) and gives rise to strong shell closure at Z = 34 for
the nuclei with N = 50 and N = 82. Therefore, our calculations predict
$^{84,116}$Se as doubly magic nuclei giving rise to shell closure at Z
= 34 similar to N = 34. Moreover, pairing energy contribution from proton and neutron is also found zero which supports the magic character of both these nuclei
$^{84,116}$Se.

Comparatively, it can be concluded that towards neutron side (N = 32) empty 1f$_{5/2}$ state lies above on 2p$_{3/2}$ and 2p$_{1/2}$ for $^{52}$Ca gives rise to N = 32 shell closure but toward proton side for Z = 32 isotopes 1f$_{5/2}$ state always lies below to 2p$_{3/2}$ and 2p$_{1/2}$ states and thus proton filling in 1f$_{5/2}$ state just after 1f$_{7/2}$ state (28+6 = 34) gives no probability of Z = 32 shell closure. It should be also mentioned here that
$^{116}$Se and $^{48}$Si both are drip-line nuclei for Z = 34 and Z
= 14 respectively and these new shell closures at Z = 34 and N = 34
are due to reorganization of single particle levels in the vicinity
of drip-lines. Again, these results have been tested for different box radius and consistency is found in the above said conclusions.

\subsection{ N = 40 and Z = 40}

The gap created by pf shell with 1g$_{9/2}$ state is responsible for
the shell closure at N = 40 or Z = 40. Therefore it would be again
interesting to investigate behaviour of pf shell, consequently respective positions of single particle
states of pf shell and 1g$_{9/2}$ state for N(Z) = 40. To
visualize this shell closure we have performed calculations for all isotones of N = 40 and isotopes
of Z = 40 with deformed RMF+BCS approach
as done in the previous subsections and found that N = 40 is spherical
for neutron rich side ranging from Z = 16 to Z = 30 having zero quadrupole deformation parameter ($\beta_{2}$ = 0), whereas Z = 40 (Zr) is
spherical for only N = 50 and N = 82. For these nuclei once again we have
applied spherical RMF+BCS approach using mass dependent pairing with NL3* parameter and some of the results are
plotted in Fig. 6.

\begin{figure}[htbp]
\begin{center}
\includegraphics[width=120mm]{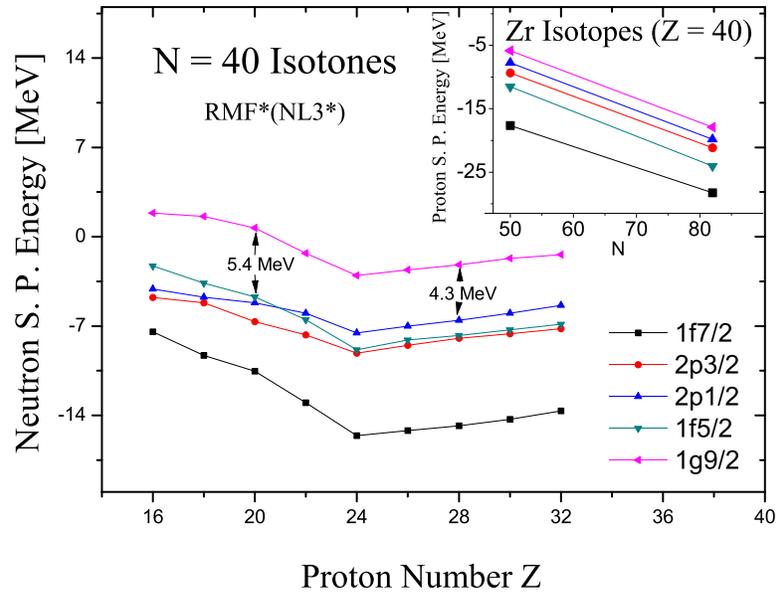}
\caption{(Colour online) Variation of
neutron single particle energy of pf shell and 1g$_{9/2}$ state with
respect to proton number Z for N = 40 isotones. The inset shows proton single particle energies for same states but for Z = 40 isotopes as a function of
Neutron number N.} \label{fig6}
\end{center}
\end{figure}
We have shown proton single
particle energy of various states for N = 40 isotones with Z =
16 - 32. From the Fig. 6, it is evident that N = 40 shell closure is
due to gap between neutron pf shell and neutron 1g$_{9/2}$ state. But the variation
in the positions of states of pf shell, this gap changes as one moves
from Z = 16 to Z = 32. For Z = 16 - 20 neutron 1f$_{5/2}$ state lies above
to the 2p$_{3/2}$ and 2p$_{1/2}$ and variation is such that gap between 1g$_{9/2}$ and 1f$_{5/2}$ states
increases from Z = 16 to Z = 20. At Z = 20 for $^{60}$Ca
1f$_{5/2}$ produces the gap with 1g$_{9/2}$ at its maximum value 5.4 MeV as can be seen
from Fig. 6. Moving towards higher Z (Z$>$20), 1f$_{5/2}$
state gets deeper because of its higher angular momentum as
compared to 2p$_{3/2}$ and 2p$_{1/2}$ states. Moving beyond Z = 20, this gap responsible for N = 40 now arises between 2p$_{1/2}$ and 1g$_{9/2}$,
and for Z = 28 it is found 4.3 MeV which results $^{68}$Ni as a candidate of doubly magicity. This doubly magic candidate $^{68}$Ni has been identified by
Broda et al. \cite{broda} and recently by  Bentley et al. \cite{bentley}.

We have also analyzed the behaviour of pf shell in Z = 40 (Zr)
isotopes. In a similar manner we found only
two isotopes which are spherical i.e. $^{90}$Zr and $^{122}$Zr as
mentioned above. For these isotopes, it is seen that there is no
reorganization of proton pf shell (1f$_{7/2}$, 1f$_{5/2}$,
2p$_{3/2}$ and 2p$_{1/2}$ states) and proton 1g$_{9/2}$ state. These
states follow the same positions while moving from $^{90}$Zr to
$^{122}$Zr as can be seen in the inset of Fig. 6. It
is found that the energy gap between pf shell and 1g$_{9/2}$ state for $^{90}$Zr and $^{122}$Zr is around 1 MeV which
is not so pronounced. Therefore for Zr isotopes, Z = 40 is not found to be
a proton shell closure dissimilar to above discussed N = 40 neutron shell closure.

Important outcome of the above study is the appearance of N = 40 shell
closure for Z = 16 to Z = 30. Among these nuclei, $^{60}$Ca and $^{68}$Ni are much
pronounced shell closures as mentioned above (gap is 5.4 MeV and 4.3 MeV respectively as mentioned also in Fig. 6). As already stated that
recently magicity of the nucleus $^{52}$Ca is confirmed by employing
the mass spectrometer of ISOLTRAP at CERN \cite{wienholtz}. So far
mass of $^{58}$Ca is known \cite{audi,wang} and therefore one may expect
the determination of mass in $^{60}$Ca in near future.
\subsection{Ground state properties of predicted magic nuclei in pf shell}

\begin{figure}[htbp]
\begin{center}
\includegraphics[width=120mm]{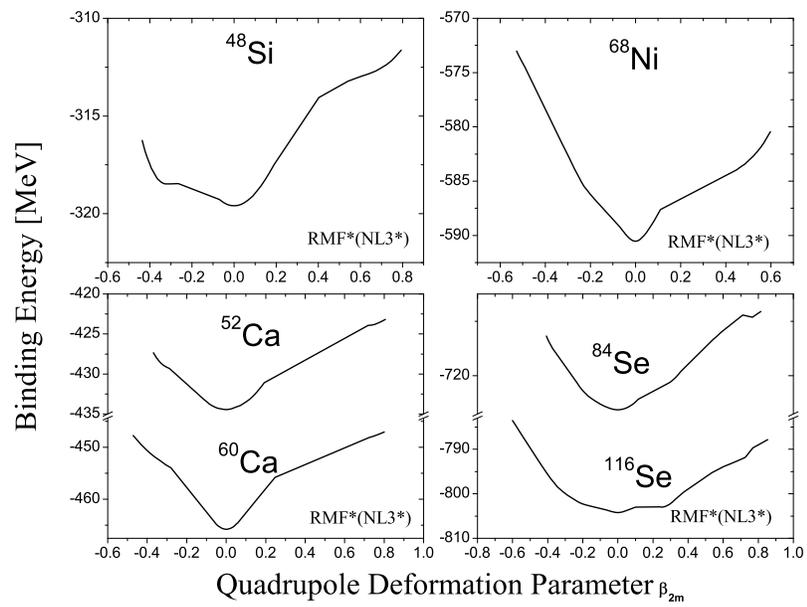}
\caption{ The potential energy surface of $^{52,60}$Ca, $^{48}$Si, $^{68}$Ni and  $^{84,116}$Se as a function of the deformation parameter $\beta_{2}$.} \label{fig7}
\end{center}
\end{figure}

With the above detailed investigations, we predict shell closure at $^{52}$Ca and strong shell closures at $^{60}$Ca, $^{48}$Si, $^{68}$Ni along with $^{84,116}$Se in pf shell. For all of these nuclei the calculations are carried out within the framework of deformed RMF approach (axially deformed configuration) and these calculations have yielded valuable results related to the ground state properties, such as binding energy, rms proton and neutron radii, two proton and two neutron separation energies, deformations etc. It will be interesting to get more insight into the shapes of these nuclei. In Fig. 7, we have plotted binding energy curves as function of quadrupole deformation parameter as obtained with constrained RMF calculations including the deformation degree of freedom for all above nuclei. From Fig. 7, it is valuable to note that all these nuclei show a spherical ground state configuration which validate all the above results interpreted  by using spherical RMF approach. A sharp minima is readily seen in $^{60}$Ca comparable to $^{52}$Ca suggesting a more strong spherical character in $^{60}$Ca. Similarly very sharp minima are also observed for $^{68}$Ni, $^{84}$Se and $^{116}$Se.
In Table 2, we present here some important ground state properties of these nuclei calculated by axially deformed RMF approach using NL3* parameter.

\begin{table}
\vspace{0.1cm} \centering \caption{Ground state properties of $^{52}$Ca, $^{60}$Ca, $^{48}$Si, $^{68}$Ni, $^{84}$Se and  $^{116}$Se }
\bigskip
\resizebox{0.88\textwidth}{!}{%
\begin{tabular}{|c|c|c|c|c|c|c|c|}
 \hline
 \multicolumn{1}{|c|}{Properties}&\multicolumn{1}{|c|}{$^{52}$Ca}&
 \multicolumn{1}{|c|}{$^{60}$Ca}&\multicolumn{1}{|c|}{$^{48}$Si}&\multicolumn{1}{|c|}{$^{68}$Ni}&\multicolumn{1}{|c|}{$^{84}$Se}&\multicolumn{1}{|c|}{$^{116}$Se}\\
 \hline
 \hline
B. E. (MeV)&434.452&465.752&319.590&590.525&726.236&804.216\\
$S_{2p}$ (MeV)&34.893&45.842&52.259&27.695&25.750&47.927\\
$S_{2n}$ (MeV)&9.90&7.049&2.445&15.770&15.374&0.767\\
$R_{c}$ (fm)&3.512&3.639&3.262&3.859&4.114&4.475\\
$R_{p}$ (fm)&3.419&3.550&3.163&3.776&4.036&4.403\\
$R_{n}$ (fm)&3.832&4.153&4.046&4.040&4.282&5.133\\
$R_{m}$ (fm)&3.681&3.962&3.810&3.933&4.184&4.930\\
$\beta_{2}$&0.00&0.00&0.00&0.00&0.00&0.00\\
Pairing Energy (MeV)&0.00&0.00&0.00&0.00&0.00&0.00\\
\hline
\end{tabular}}
\label{tab15}
\end{table}

Moreover, guided from the predicted shell closures and for systematic comparison with experimental data we have also calculated ground state properties of chains of isotopes of Si, Ca, Ni and Se isotopes using axially deformed RMF approach with NL3* parameter. In Fig. 8, we have plotted  two neutron separation energy S$_{2n}$ for chains of isotopes of Si, Ca, Ni and Se nuclei. To demonstrate the validity of RMF calculations, we have also compared our results of two neutron separation energy with one of the popular nonrelativistic approach viz. Skyrme-Hartree-Fock method with the HFB-17 functional given by Goriely et al. \cite{goriely1}. From Fig. 8 it is evident that results of RMF calculations are in fairly good agreement with available experimental data and the results of Skyrme-Hartree-Fock method for all three chains of isotopes. For Si isotopes, the two neutron separation energy becomes negative after N = 34 with RMF calculations and concludes that two neutron drip line for Si isotopes is N = 34. Interestingly this nucleus of Si with N = 34 ($^{48}$Si) is one of our predicted doubly magic nucleus as mentioned earlier and leads to an example of drip line doubly magic nucleus with new neutron magic number N = 34. However, with HFB calculations \cite{goriely1} this nucleus $^{48}$Si is little unbound with 0.46 MeV resulting two neutron drip line at N = 32 for Si isotopes. In a similar way drip lines for Ca, Ni and Se isotopes with RMF calculations are found at N = 46, N = 70 and at N = 82 respectively. Therefore, our predicted doubly magic nucleus $^{60}$Ca is near to drip line of Ca and once again another predicted doubly magic nucleus $^{116}$Se represents an interesting example of drip line doubly magic nucleus with new proton magic number Z = 34. From HFB calculations the drip lines are found at little variance N = 48, N = 64 and N = 80 for Ca, Ni and Se isotopes respectively. A sharp decrease in the value of S$_{2n}$ just after magic number can be seen from the Fig. 8. This sharp fall lies after $^{48}$Si, $^{52}$Ca, $^{60}$Ca, $^{68}$Ni, $^{84}$Se and $^{116}$Se affirming our conclusions. Other properties of these isotopic chains and similar systematic calculations for chains of isotones of N = 32, 34 and 40 (not shown here) are found with excellent agreement with available experimental data \cite{audi} along with HFB calculations \cite{goriely1} and over again fortify our predictions.

\begin{figure}[htbp]
\begin{center}
\includegraphics[width=120mm]{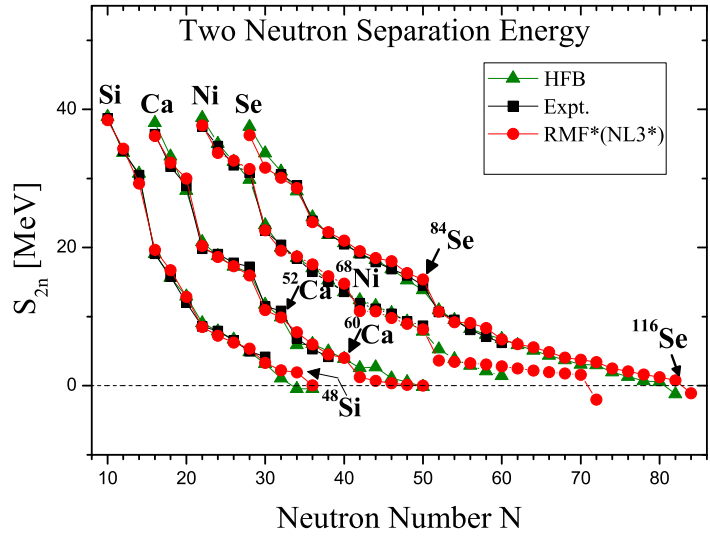}
\caption{(Colour online) The two neutron separation energy for Si, Ca, Ni and Se isotopes are obtained with axially deformed RMF approach and compared with the Skyrme-Hartree-Fock
calculations \cite{goriely1}. Figure also depicts the available experimental data \cite{audi} for the purpose of comparison.} \label{fig8}
\end{center}
\end{figure}

\section{Summary}
In the present investigation we have employed the relativistic
mean-field plus BCS (RMF+BCS) approach \cite{saxena1,singh} to
study shell closures N(Z) = 32, 34 and 40 in neutron and proton pf shell extensively.
For our study, the RMF calculations for all the nuclei considered here have been firstly carried out assuming a deformed
shape (deformed RMF) and then the situation where nuclei are ascertained exhibiting very small or almost no deformation has been beneficially utilized
by employing spherical RMF approach. For our calculation, mass dependent pairing is found more suitable and effective over constant pairing to cover a large mass region. The main body of the results of our  calculations includes single particle spectra, two proton and two neutron separation
energies and other ground state properties for complete isotonic and isotopic chain. For the description we have applied mainly NL3* \cite{nl3star} parameter and used various other popular force parameters viz. TMA \cite{suga-tma}, NL-SH \cite{ring3-nlsh}, NL3 \cite{lalanl3} and PK1 \cite{pk1} to testify our results and outcomes.

One of the prime reason of this study is to investigate neutron and proton single particle states of pf shell and consequently appearance of new shell closures. The main force behind this study is recent experimental observation showing new magicity at N = 32 and 34 \cite{gade,wienholtz,stepp,rosenbusch}. We have found through single particle energies and separation energies that N = 32 behaves like a shell closure in $^{52}$Ca which is in accord of recent observation, whereas N = 34 is not showing shell closure in $^{54}$Ca. But, N = 34 is found to possess more strong shell closure in proton deficient nucleus $^{48}$Si which is an example of doubly magic drip-line nucleus and potential candidate for future experiments in the same mass region as that of $^{52}$Ca. Towards proton side Z = 32 is not found to exhibit magic character but Z = 34 becomes closed and resulting doubly magic nuclei $^{84,116}$Se out of which $^{116}$Se is again an example of doubly magic drip-line nucleus. In addition to this, we have also focused on shell closure N(Z) = 40 and came up that N = 40 shows shell closure towards proton deficient side resulting doubly magicity in $^{60}$Ca and $^{68}$Ni. Out of these nuclei the gap between neutron pf shell and next neutron 1g$_{9/2}$ state responsible of N = 40 shell closure, is maximum for $^{60}$Ca. Our further study of two neutron shell gap for Ca isotopes also confirms that $^{60}$Ca is possibly another important doubly magic nucleus near drip-line next to $^{52}$Ca for future experiments.

\section*{Aknowledgements}
Authors would like to thank Prof. H. L. Yadav, Banaras Hindu University,
Varanasi, INDIA for his kind guidance and continuous support. The authors are indebted to Prof. L. S. Geng, Beihang University, China
for valuable correspondence. One of the authors (G. Saxena)
gratefully acknowledges the support provided by Science and Engineering Research Board (DST),
Govt. of India under the young scientist project YSS/2015/000952.

\bibliographystyle{model1a-num-names}

\end{document}